\documentclass[12pt]{article}
\usepackage[margin=1in]{geometry}
\usepackage[utf8]{inputenc}
\usepackage{amsmath}
\usepackage{graphicx}
\usepackage{subcaption}
\usepackage{hyperref}
\usepackage{listings}
\usepackage[HTML]{xcolor}
\usepackage{tikz}
\usepackage[utf8]{inputenc}
\usepackage[T1]{fontenc}
\usepackage[font=small,labelfont=bf,tableposition=top]{caption}
\usetikzlibrary{shapes.geometric, arrows.meta, positioning}

\tikzset{
    block/.style = {rectangle, rounded corners, minimum width=2.8cm, minimum height=0.8cm,
                    text centered, draw=black, fill=blue!20, font=\normalsize},
    decision/.style = {diamond, aspect=2, text centered, draw=black, fill=green!20, font=\normalsize},
    arrow/.style = {thick, ->, >=Stealth}
}

\title{Integration of Variational Quantum Algorithms into Atomistic Simulation Workflows}
\author{Wilke Dononelli$^{1,*}$}
\date{}

\begin{document}
\maketitle

{\footnotesize
\begingroup
\raggedright
\hangindent=0.7em
\hangafter=1
$^1$ Institute for Physical and Theoretical Chemistry, Bremen Center for Computational Materials Science University of Bremen and MAPEX Center for Materials and Processes, Leobener Str.~6, D-28359~Bremen, Germany. 
$*$ correspond to \href{mailto:wido@uni-bremen.de}{wido@uni-bremen.de}
\endgroup}

\begin{abstract}
In this work, we present the integration of Qiskit Nature’s quantum chemistry solvers into the Atomic Simulation Environment (ASE), enabling hybrid quantum–classical workflows for force-driven atomistic simulations. This coupling allows the use of the Variational Quantum Eigensolver (VQE) and its adaptive variant (ADAPT–VQE) not only for ground-state energy calculations, but also for geometry optimisation, vibrational frequency analysis, strain evaluation, and molecular dynamics, all managed through ASE’s calculator interface. By applying ADAPT–VQE to multi-electron systems such as BeH$_2$, we obtain vibrational and structural properties in close agreement with high-level classical CCSD calculations within the same minimal basis. These results demonstrate that adaptive variational quantum algorithms can deliver stable and chemically meaningful forces within an atomistic modelling workflow, enabling downstream applications such as molecular dynamics and active-learning–accelerated simulations.
\end{abstract}
\begin{figure}[htb]
  \centering
  \includegraphics[width=0.4\linewidth]{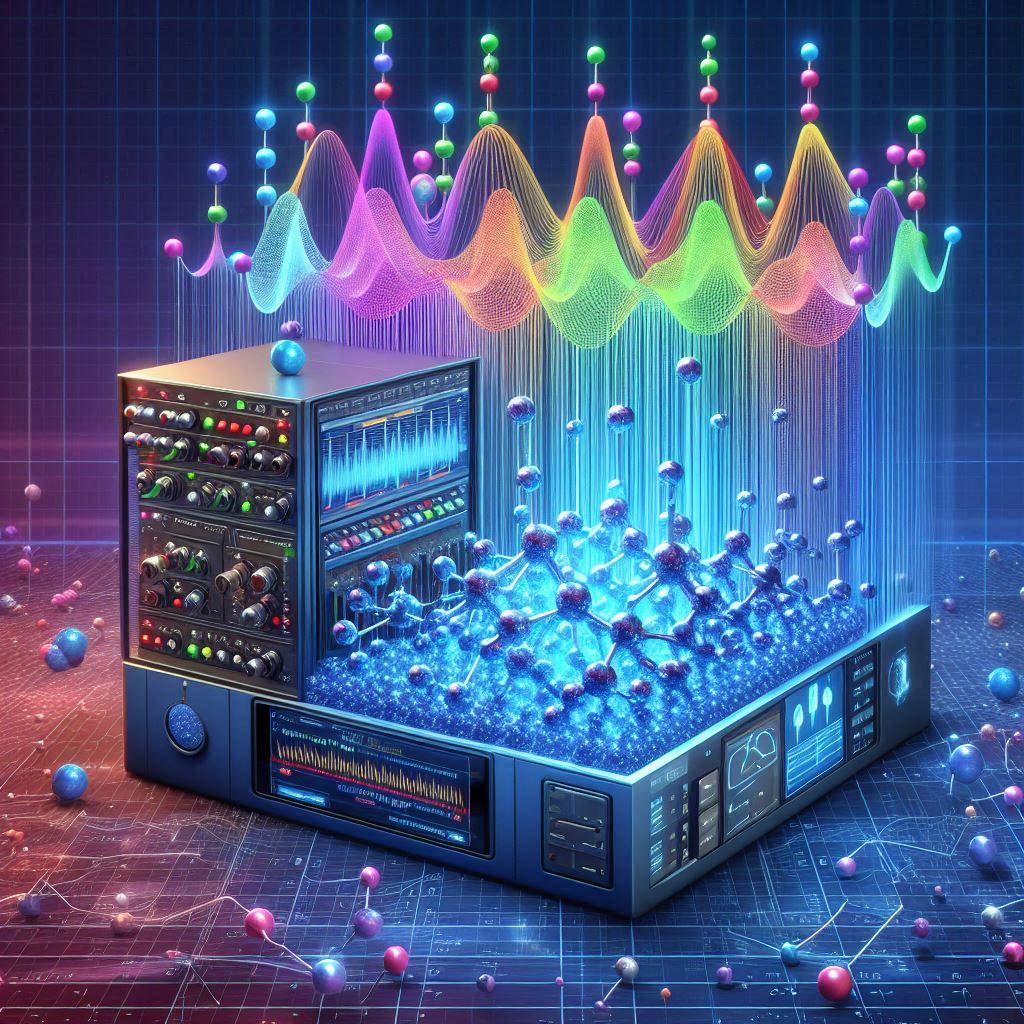}
  \label{fig:designer}
\end{figure}
\section{Introduction}
Advances in quantum algorithms have generated intense interest in their potential to complement classical approaches to electronic structure problems, particularly in regimes where the steep scaling of high-level methods becomes prohibitive. In atomistic modelling, wide-ranging applications such as catalysis design, battery material discovery, and molecular spectroscopy rely on accurate solutions to the electronic Schrödinger equation~\cite{norskov2011dft,ceder2013stuff,chan2011dmrg,cramer2004essentials,marx2009ab,curtarolo2013high}. High-level classical methods such as coupled cluster with singles and doubles (CCSD) or with perturbative triples [CCSD(T)], often considered the gold standard of quantum chemistry~\cite{bartlett2007coupled}, achieve near-chemical accuracy but scale steeply with system size, typically as $\mathcal{O}(N^{6})$ and $\mathcal{O}(N^{7})$ respectively. This scaling sharply limits application to small molecules in practice.

Early quantum algorithms for chemistry, such as quantum phase estimation (QPE)\cite{aspuru2005simulated,lloyd1996universal}, promised exact eigenvalues but required circuit depths far beyond the capabilities of noisy intermediate-scale quantum (NISQ) devices\cite{omalley2016scalable}. The variational quantum eigensolver (VQE)\cite{peruzzo2014variational} emerged as a hybrid quantum–classical strategy that dramatically reduces circuit depth by combining state preparation and measurement on a quantum processor with classical optimisation of a parametrised wavefunction ansatz. Chemically motivated ansätze such as unitary coupled cluster with singles and doubles (UCCSD)\cite{taube2006new,bartlett2007coupled} have been widely adopted, alongside hardware-efficient forms~\cite{kandala2017hardware} and low‑depth generalisations such as k‑UpCCGSD~\cite{lee2019generalized}. Further refinements like the quantum subspace expansion~\cite{colless2018computation} and qubit coupled cluster~\cite{ryabinkin2018qubit} balance expressivity against circuit resources.

In practice, however, many demonstrations have been confined to so‑called ``toy systems''—minimal basis calculations on very small molecules—where the quantum measurement is exact and the parameter convergence is trivial. Such examples validate implementations but fall short of showcasing chemistry-level utility. For workflow-oriented applications, accurate and *stable* forces and vibrational properties in multi-electron systems must be obtained under the constraints of finite sampling (shots), noisy measurement, and challenging classical optimisation. In particular, molecular dynamics and structure exploration place stringent demands on the consistency of forces across many geometries, making force robustness, rather than single-point energy accuracy alone, a central bottleneck for quantum algorithms in atomistic modelling. 
For larger systems, these challenges are further exacerbated by the substantial measurement overhead required for expectation-value–based variational algorithms~\cite{Gonthier2022} and by optimization pathologies such as barren plateaus in high-dimensional parameter landscapes~\cite{McClean2018}.

Recent algorithmic advances in the NISQ era have focussed on reducing required quantum resources and improving convergence for realistic quantum chemical problems. Among the promising tools are adaptive ansatz constructions such as the Adaptive Derivative-Assembled Pseudo-Trotter VQE (ADAPT–VQE)\cite{ADAPT1,ADAPT2}, which iteratively builds the ansatz by selecting the most important excitation operators based on energy gradients. This approach yields compact, system-specific circuits which have demonstrated improved accuracy and convergence, especially in strongly correlated regimes\cite{ADAPT1,ADAPT2,magnusson2024tcavqite,grimsley2023adaptive}. Parallel efforts have aimed at further circuit compression and more accurate wavefunction representations, for example, by using transcorrelated (TC) Hamiltonians~\cite{magnusson2024tcavqite}, which incorporate explicit electron-electron correlation via Jastrow factor transformations~\cite{boys1969TC,hirschfelder1963TC,dobrautz2019compact}. Recent work~\cite{magnusson2024tcavqite} shows that combining the TC approach with adaptive quantum algorithms such as AVQITE and VQE can significantly reduce quantum circuit depth and width, enhance convergence, and improve noise resilience, thereby facilitating their use in iterative atomistic workflows that require repeated and reliable force evaluations.

At the software ecosystem level, Qiskit Nature~\cite{qiskit2023nature}, part of the Qiskit SDK~\cite{Qiskit}, provides robust implementations of VQE, ADAPT–VQE, and related algorithms, interfacing with molecular integrals from classical quantum chemistry packages such as PySCF. 

However, its current scope is limited to isolated quantum chemistry calculations; it does not directly offer the broader atomistic modelling functionality—geometry optimisation, vibrational analysis, and molecular dynamics—that underpin materials and molecular simulation workflows. Related efforts have explored interfacing variational quantum algorithms with atomistic environments, primarily focusing on single-point energies or force evaluations, but without addressing the demands of long-time dynamical simulations.
Previous studies have explored interfacing variational quantum eigensolvers with atomistic simulation environments, including force evaluations within ASE frameworks~\cite{Sarkar2022,Nishida2025}. In contrast, the present work focuses on enabling computationally tractable molecular dynamics by combining ADAPT‑VQE forces with on‑the‑fly active learning. While Nishida~\cite{Nishida2025} proposes a quantum circuit learning approach that replaces variational optimization with a pretrained feedforward model, the present work instead retains variational quantum algorithms and addresses their computational cost at the workflow level through classical active learning.

Here, we integrate Qiskit Nature’s quantum–classical solvers into the widely used Atomic Simulation Environment (ASE)~\cite{ASE} as an enabling layer for force-driven atomistic workflows based on variational quantum algorithms. This allows the electronic structure step in ASE’s geometry optimisation, vibrational frequency analysis, strain evaluation, and molecular dynamics to be performed using VQE or ADAPT–VQE, with results from a simulator returned to ASE’s property interface. The integration permits drop-in replacement of classical electronic structure calculators with quantum counterparts, while preserving the surrounding atomistic workflow. Previous studies have explored interfacing variational quantum eigensolvers with atomistic simulation environments, including force evaluations within ASE frameworks~\cite{Sarkar2022,Nishida2025}. These efforts primarily focused on proof‑of‑concept geometry optimizations or single‑point force evaluations, without addressing the challenges posed by long‑time molecular dynamics or force‑driven workflows. In contrast, in this present work we emphasize the use of adaptive variational quantum algorithms in combination with on‑the‑fly active learning to enable computationally tractable molecular dynamics simulations.

By combining the gernal workflow integration with advanced quantum algorithms such as ADAPT–VQE, we aim to demonstrate that stable and chemically meaningful forces can be obtained for multi‑electron, polyatomic systems within a hybrid quantum–classical framework. This, in turn, enables downstream applications including geometry optimisation, vibrational analysis, and molecular dynamics, rather than limiting quantum algorithms to isolated single-point calculations. The following sections detail the implementation, benchmark results against classical CCSD, and representative applications ranging from simple diatomics to triatomics and force-driven dynamical simulations. Note however, that we still perform all calculations in this study in the limit of using a minimal basis set. However, the finding should be transferable to the usage of bigger basis set sizes and therefore, the here presented workflows are ready to be transferred to bigger system sizes in the future. In this context, embedding and partitioning approaches, in which only chemically relevant subsystems are treated quantum mechanically while the environment is handled classically, provide a promising route toward extending such workflows to larger and more complex systems~\cite{Vorwerk2022}, although they are out of the scope of the here underlying study. 

Although the code is able to run in a hybrid mode on real NISQ devices, here we focus on simulated devices and on the VQE and ADAPT–VQE algorithms. This choice allows systematic investigation of convergence behaviour and force stability relevant to dynamical simulations, while future work could extend the framework to more advanced algorithms and real-device executions.

The code is publicly available at \url{https://github.com/thequantumchemist/ase_quantum_vqe/} and can be used and modified.
\section{Computational Methods}
The core of this work is a custom ASE \texttt{Calculator} that wraps a Qiskit Nature electronic structure calculation within ASE's~\cite{ASE} standard interface. In this implementation, the geometry described by the ASE \texttt{Atoms} object is passed to the \texttt{PySCFDriver}~\cite{pyscf} in Qiskit Nature~\cite{qiskit2023nature}, which computes the one- and two-electron integrals for the system using the specified Gaussian basis set. The resulting second-quantised fermionic Hamiltonian is mapped to qubit operators via the Jordan--Wigner mapping~\cite{jordan1928}.

A variational quantum eigensolver (VQE) is then prepared using the unitary coupled cluster ansatz with single and double excitations (UCCSD)~\cite{taube2006new,bartlett2007coupled}, combined with a Hartree--Fock reference state. In addition to fixed UCCSD ansätze, the calculator supports adaptive ansatz construction via ADAPT--VQE, in which excitation operators are iteratively selected based on energy gradients and appended to the variational circuit. The variational parameters are optimised by a classical optimizer (SLSQP in this work if not stated otherwise) with the cost function given by expectation values obtained from the Qiskit \texttt{Estimator}. The calculator is flexible in that it can employ either \texttt{qiskit-aer}'s statevector simulator \textit{aer} or connect to an IBMQ backend via the Qiskit Runtime service, allowing tokens and backend selection to be specified at runtime. A detailed workflow is provided in Fig. \ref{fig:vqe_workflow}.

\tikzstyle{block} = [rectangle, rounded corners,
    minimum width=2.8cm, minimum height=0.8cm,
    text centered, draw=black, fill=blue!20, font=\normalsize]
\tikzstyle{decision} = [diamond, aspect=2, text centered,
    draw=black, fill=green!20, font=\normalsize]
\tikzstyle{arrow} = [thick,->,>=stealth]

\begin{figure}[ht]
\centering
\resizebox{0.45\linewidth}{!}{%
\begin{tikzpicture}[node distance=1.3cm]

\node (start) [block] {ASE: Geometry};
\node (attach) [block, below of=start] {Attach VQE Calc};
\node (driver) [block, below of=attach] {PySCFDriver: integrals};
\node (mapper) [block, below of=driver] {Active space + JW map};
\node (ansatz) [block, below of=mapper] {Ansatz: HF $\to$ UCCSD};
\node (choice) [decision, below of=ansatz, yshift=-0.3cm] {Backend};
\node (aer) [block, left of=choice, xshift=-4cm] {Aer Simulator};
\node (ibmq) [block, right of=choice, xshift=4cm] {IBM Q Runtime};
\node (vqe) [block, below of=choice, yshift=-0.7cm] {Run VQE / ADAPT-VQE};
\node (results) [block, below of=vqe] {Energy, Forces, Dipole};
\node (ase) [block, below of=results] {ASE tasks e.g. geometry optimization, frequencies, strain, MD, ...};

\draw [arrow] (start) -- (attach);
\draw [arrow] (attach) -- (driver);
\draw [arrow] (driver) -- (mapper);
\draw [arrow] (mapper) -- (ansatz);
\draw [arrow] (ansatz) -- (choice);
\draw [arrow] (choice.west) -- (aer.east);
\draw [arrow] (choice.east) -- (ibmq.west);
\draw [arrow] (aer.south) |- (vqe.west);
\draw [arrow] (ibmq.south) |- (vqe.east);
\draw [arrow] (vqe) -- (results);
\draw [arrow] (results) -- (ase);

\end{tikzpicture}
}
\caption{Compact workflow of ASE--QiskitVQECalculator: ASE passes geometry to Qiskit Nature for VQE/ADAPT-VQE on a selected backend, returning energies, forces, and properties to ASE.}
\label{fig:vqe_workflow}
\end{figure}
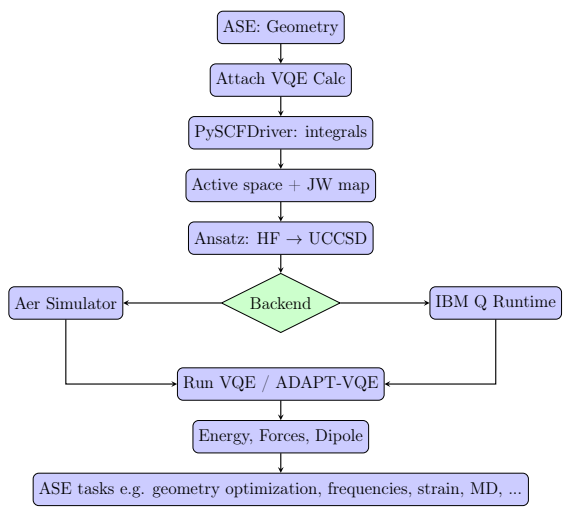

The total energy computed in atomic units is converted to eV for ASE. Forces are determined numerically by central finite differences, displacing each atomic coordinate by a small step and recomputing the energy. To ensure force consistency across geometry optimisation, vibrational analysis, and molecular dynamics, all displaced geometries in a force evaluation reuse identical electronic structure settings and convergence thresholds, thereby minimising numerical noise between successive force calls. Dipole moments are extracted in atomic units from Qiskit Nature's internal auxiliary operators and converted to Debye for compatibility with ASE's \texttt{dipole} property, enabling direct use of the \texttt{Infrared} module for IR spectra.\\

The integration in ASE allows for pre-defined default settings. Below we show two sample input scripts: a simple example illustrating black-box geometry optimisation of an H$_2$ molecule, and a more advanced example demonstrating global optimisation of H$_3^+$ followed by local optimisation, vibrational analysis, and strain evaluation.\\
~\\
~\\
Minimal example of geometry optimisation of an H$_2$ molecule:\\
\begin{table}[ht]
\centering
\begin{tabular}{|p{0.65\linewidth}|}
\hline
\begin{scriptsize}
\begin{verbatim}
from ase.build import molecule
from ase.optimize.bfgs import BFGS
from ase_quantum_vqe.qiskit_vqe_calculator import QiskitVQECalculator

atoms = molecule('H2')
atoms.calc = QiskitVQECalculator()

opt = BFGS(atoms, trajectory='opt.traj')
opt.run()
\end{verbatim}
\end{scriptsize}
\\ \hline
\end{tabular}
\end{table}
~\\
The algorithm uses default parameters, such as execution on the \textit{aer} simulator rather than a real backend. More advanced parameters can be specified as needed. An example illustrating a frequency calculation and strain analysis for an H$_3^+$ molecule is shown below:\\
\begin{table}[htbp]
\centering
\begin{tabular}{|p{0.65\linewidth}|}
\hline
{\scriptsize
\begin{verbatim}
from ase import Atoms
from ase.optimize.bfgslinesearch import BFGSLineSearch as BFGS
from ase_quantum_vqe.qiskit_vqe_calculator import QiskitVQECalculator
from ase.vibrations import Vibrations
from strainjedi.jedi import Jedi
from ase.optimize.minimahopping import MinimaHopping
from ase_quantum_vqe.qiskit_vqe_calculator.utils import /
           random_positions_with_min_distance as random_pos

# create molecule
positions = random_pos(n_atoms=3, min_dist=0.6, box_size=3.0)
atoms = Atoms("H3", positions=positions)

# set the calculator
calc = QiskitVQECalculator(
    basis='sto3g',
    backend='aer',
    n_jobs=18,
    charge=1,
    spin=0,
    delta=0.01,
    shots=2000,
    resilience_level=0,
    optimizer_name='COBYLA',
    coreorb=0,
    maxiter=100)

atoms.calc = calc

# global optimisation
hop = MinimaHopping(atoms, Ediff0=2.5, T0=4000.0)
hop(totalsteps=50)

# local optimisation
dyn = BFGS(atoms, trajectory='h3_vqe.traj')
dyn.run(fmax=0.005)

# vibrational analysis
vib = Vibrations(atoms, name='vibvqe', nfree=2)
vib.run()
vib.summary()
hessian = vib.get_vibrations()

# displaced structure
atomsb = atoms.copy()
atomsb.positions[1][2] += 0.1
atomsb.calc = calc
atomsb.get_potential_energy()

# strain analysis
j = Jedi(atoms, atomsb, hessian)
j.set_bond_params(covf=2.0, vdwf=0.9)
j.run()
j.vmd_gen()
\end{verbatim}
}
\\ \hline
\end{tabular}
\end{table}
In the example above, strain is analysed using the Judgement of Energy DIstribution method (\textit{JEDI}) as implemented in ASE~\cite{stauchjedi,wangjedi}. Prior to vibrational or \textit{JEDI} analyses, structural optimisations were performed using the BFGS~\cite{BFGS_press1992numerical,BFGS_nocedal1999numerical} (Broyden--Fletcher--Goldfarb--Shanno) algorithm as implemented in ASE, with a convergence criterion of 0.005~eV/\AA\ for the maximum force. Molecular dynamics simulations were conducted using ASE's Langevin dynamics~\cite{langevin_allen1987computer}. Because direct VQE force evaluations are computationally expensive for extended molecular dynamics, on-the-fly active learning was employed using the FALCON framework~\cite{felisfalcon}. In this approach, a Gaussian-process surrogate model is trained during the simulation and selectively queried with VQE calculations, enabling stable and computationally tractable molecular dynamics trajectories while retaining quantum-mechanical force fidelity.\\

If not stated otherwise, all calculations employ the \texttt{STO-3G} basis set. While this basis is insufficient for chemically accurate absolute predictions, it allows reproducible benchmarking of quantum and classical methods within a fixed representation and keeps circuit sizes compatible with current simulators.\\
The code is publicly available at \url{https://github.com/thequantumchemist/ase_quantum_vqe/} and can be used and modified. 
\section{Results}
\subsection{Proof of principle}
A key goal of this work is to move beyond proof-of-concept singlepoint calculations on one- and two-electron systems and demonstrate that hybrid quantum–classical algorithms, particularly the Variational Quantum Eigensolver (VQE) and especially its adaptive variant (ADAPT–VQE), can deliver chemically relevant properties within an atomistic modelling workflow. Unless otherwise stated, all electronic structure runs were performed in the minimal \texttt{STO-3G} Gaussian basis to keep the qubit counts and circuit dimensions compatible with current simulators and early-stage quantum hardware. It is important to note that this basis is significantly smaller than what is required for spectroscopic accuracy: absolute vibrational frequencies are systematically overestimated compared to experiment due to basis set incompleteness. However, within this fixed basis, consistent comparisons between conventional CCSD, VQE–CCSD, and ADAPT–VQE–CCSD directly reflect the intrinsic behavior of the quantum algorithms.

\subsubsection{H\texorpdfstring{$_2$}{2}}

As a baseline, we applied our ASE--Qiskit VQE calculator to the simplest diatomic molecule, H$_2$.
Starting from an initial bond length of 0.90~\AA, a geometry optimization using the UCCSD ansatz on
the \texttt{qiskit-aer} statevector simulator converged to an equilibrium distance of 0.735~\AA,
with a computed total energy of $-30.948$~eV. A classical exact calculation (directly performed in
PYSCF) within the same \texttt{STO-3G} basis yielded identical bond length and energy to within an
error of $10^{-5}$~eV, verifying the correctness of the implementation. Harmonic vibrational analysis
gave a stretching frequency of 5002~cm$^{-1}$ from VQE, compared to 5000~cm$^{-1}$ classically---a
relative deviation of less than 0.04\%. However, as expected, using a proper basis set such as
\textit{cc-pVQZ} lengthens the bond to 0.742~\AA\ and subsequently lowers the frequency to
4409~cm$^{-1}$. This indicates that the dominant error source in current electronic structure
calculations on quantum devices is not the algorithm but the basis truncation.

\subsubsection{F\texorpdfstring{$_2$}{2}}

For a more demanding case, we examined the molecular fluorine, F$_2$, which in principle has a substantially larger correlation space. However, as simple test a frozen core approximation for the correlation part of the calculation was used and only one doubly occupied orbital was kept active. Using CCSD with a \texttt{STO-3G} basis in PYSCF as our classical reference, we obtained a stretching frequency of 1673.2~cm$^{-1}$ and a total energy of –198.365~eV. In contrast, a straightforward VQE–CCSD with fixed UCCSD ansatz yielded a frequency of 7317.2~cm$^{-1}$, reflecting a large overestimation due to the difficulty of converging the variational parameters in the high-dimensional ansatz and numerical noise in finite-difference forces. This results was surprising, since the number of valence orbitals is similar than the ones in the H$_2$ example. \\
In the VQECalculator, forces are finite differences of VQE energies at slightly displaced geometries. If the energy of each displaced geometry is not fully converged with respect to the variational parameters before evaluation, the finite difference will pick up large numerical errors. Any variation in optimization quality between $+\delta$ and $–\delta$ displacement worsens the noise. Many VQE calculator implementations currently start the variational parameters from scratch for each geometry (and for each displacement in a force calculation). This \textit{cold start} greatly slows convergence. In classical geometry optimizations we reuse the MO guess from the previous geometry (wavefunction continuation). Nevertheless, an implementation of such \textit{a warm start} from converged parmeters, reduced the numerical noise only sligthly. In addtion, the classical optimiser in VQE (e.g. SLSQP, COBYLA) can start far away from the true minimum in this high dimensional space, so the initial state may have large energy and bad forces. Nevertheless, we tested three different optimizers and non of them could erase the numerical noise completely. However, even on simulators, if you use sampling (shots), the measurement noise adds uncertainty, making the numerical gradient less stable. On real devices, noise and drift magnify this effect.\\

Employing the ADAPT–VQE procedure markedly improved the results. After only 10 adaptive iterations, the frequency dropped to 1675.6~cm$^{-1}$, and by 100 iterations, the value was 1673.2~cm$^{-1}$, indistinguishable from the classical CCSD reference within the resolution of the vibrational calculation. The relative error was reduced from $\approx 337\%$ for the fixed-ansatz VQE to below $0.1\%$ for ADAPT–VQE, demonstrating that beyond trivial systems, an adaptive ansatz is essential for extracting accurate force constants and vibrational spectra on near-term hardware. Building up on this finding, the default method in the Calculator has been set to ADAPT-VQE.

\subsubsection{BeH\texorpdfstring{$_2$}{2}}

Beryllium dihydride further increases the electron count and spatial degrees of freedom, posing a stronger challenge for parameter optimization. Classical CCSD again using a \texttt{STO-3G} basis predicts three vibrational modes at 780.4, 2303.5, and 2573.8~cm$^{-1}$. The VQE–CCSD produces again highly scattered results, with stretches and bends shifted by hundreds of cm$^{-1}$, e.g. 1026.5, 1444.4, and 2303.2~cm$^{-1}$, reflecting unstable numerical gradients under finite displacements (see Tab. \ref{tab:beh2}).
\begin{table}[bth!]
\begin{tabular}{llll}\hline
Method            & v1     & v2     & v3      \\\hline
conventional CCSD & 780.4  & 2303.5 & 2573.8  \\
VQE CCSD          & 1026.5 & 1444.4 & 2303.2  \\
ADAPT-VQE CCSD    & 725.7  & 2303.5 & 2580.3\\ \hline
\end{tabular}
\caption{Comparison of the vibrational modes of BeH$_2$ from a classical quantum chemical calculation to VQE and ADAPT-VQE results}
\label{tab:beh2}
\end{table}

ADAPT–VQE with carefully tightened convergence thresholds, increased shot counts, and longer classical optimizer runs stabilizes the spectrum, yielding 725.7, 2303.5, and 2580.3~cm$^{-1}$. The largest deviation from the CCSD benchmark is now $\approx 7\%$ for the lowest-frequency mode, while high-frequency modes are recovered within $0.3\%$. These findings make clear that while bending modes in multi-atom systems remain sensitive to residual ansatz incompleteness and noise, the adaptive approach achieves a step-change in agreement to the calssically obtained result compared to fixed-ansatz VQE.

\subsection{Advanced applications}

\subsubsection{H\texorpdfstring{$_3^+$}{3+}: Global Optimization, Vibrations, and Strain Analysis}

\begin{figure}[!hbt]
  \centering
  \begin{minipage}[t]{0.35\textwidth}
    \vspace{0pt} 
    \includegraphics[width=\linewidth]{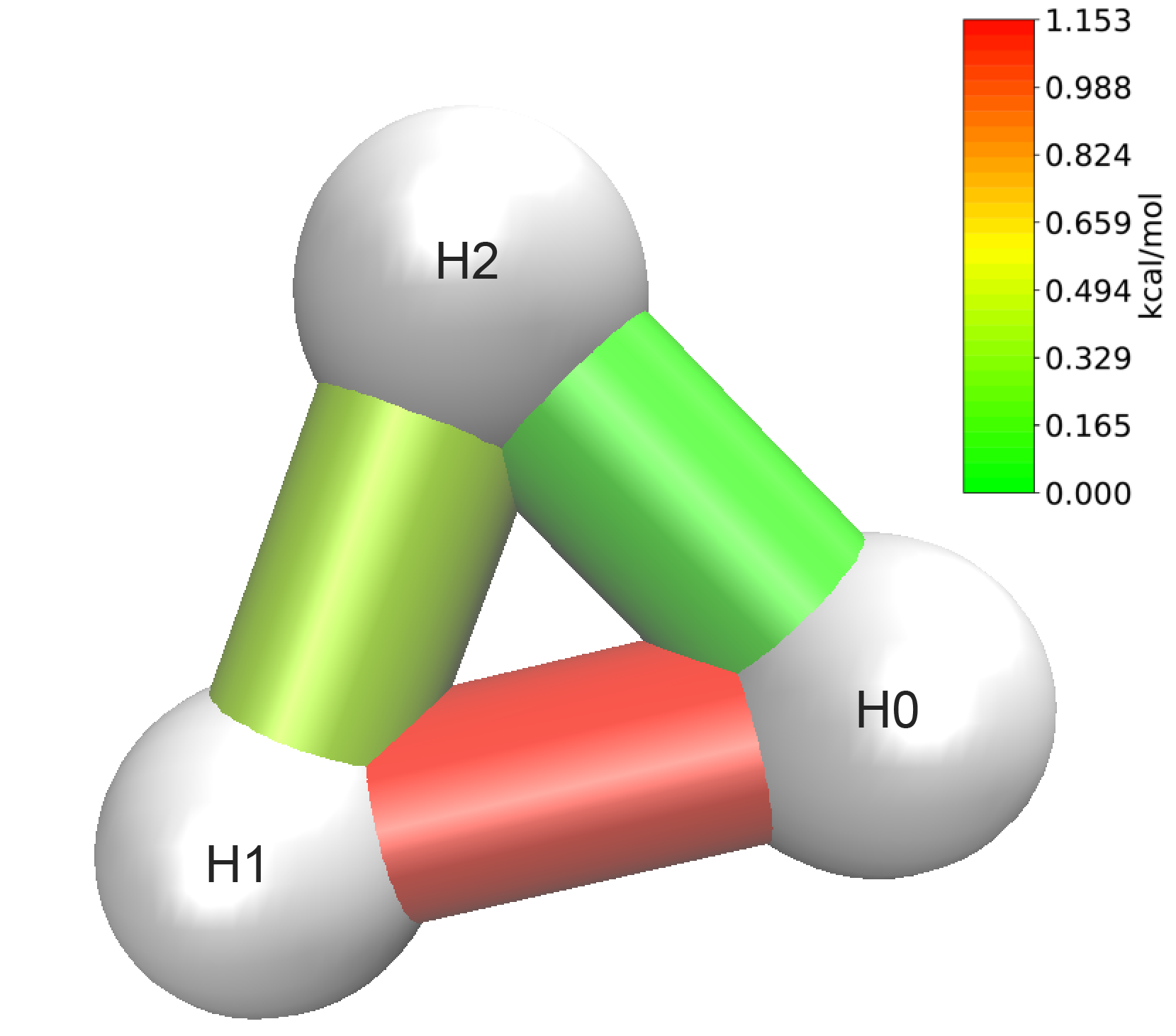}
  \end{minipage}%
  \hspace{0.04\textwidth}
  \begin{minipage}[t]{0.60\textwidth}
    \vspace{0pt} 
    \begin{footnotesize}
    \begin{tabular}{llll}\hline
             &          & conv. CCSD & ADAPT-VQE \\ \hline
      vibrational modes  & v1       & 2113.9            & 2125.9    \\
      (cm$^{-1}$)        & v2       & 2118.7            & 2127.1    \\
             & v3       & 3446.0            & 3446.1    \\
             &          &                   &           \\ \hline
      Strain (kcal/mol)        & indices  &                   &           \\ \hline
      bond                     & H0 H1    & 1.0954471         & 1.0912963 \\
      bond                     & H0 H2    & 0.0000000         & 0.0000000 \\
      bond                     & H1 H2    & 0.2352117         & 0.2350237 \\
      angle                    & H0 H1 H2 & 0.1084832         & 0.1075360 \\
      angle                    & H0 H2 H1 & 0.1704137         & 0.1687706 \\
      angle                    & H1 H0 H2 & 0.0069441         & 0.0068524 \\ \hline
			\label{tab:h3+}
    \end{tabular}
    \end{footnotesize}
  \end{minipage}
  \caption{Vibrational modes and strain energies of H$_3^+$}
	\label{fig:h3+}
\end{figure}

To test a complete hybrid workflow beyond spectral properties, we applied global minimum searching via ASE’s minima hopping algorithm to the triatomic cation H$_3^+$. The search identified the known equilateral minimum structure. Subsequent local geometry optimization and vibrational analysis with ADAPT–VQE yielded modes at 2125.9, 2127.1, and 3446.1~cm$^{-1}$, in good agreement with the classical CCSD results of 2113.9, 2118.7, and 3446.0~cm$^{-1}$, respectively (c.f. table in fig. \ref{fig:h3+}).
In a next step, the mechanical strain was tested via the JEDI method. Displacing one hydrogen
along the molecular plane introduced a total strain energy of 1.091~$\mathrm{kcal\,mol^{-1}}$
(ADAPT-VQE), compared to 1.095~$\mathrm{kcal\,mol^{-1}}$ for CCSD. Bond-localized strain energies
matched to within 0.004~$\mathrm{kcal\,mol^{-1}}$, showing the ADAPT-VQE ansatz reproduces subtle
energy changes associated with small geometry distortions.

\subsection{Molecular Dynamics enhanced by on-the-fly trained Machine Learning Potential using VQE Energies and Forces}
As a final demonstration, we tested whether the numerically evaluated ADAPT–VQE forces were robust enough for extended molecular dynamics (MD) simulations. Before addressing molecular dynamics of multi-electron polyatomic systems, we first assess the stability and efficiency of force-driven molecular dynamics in the simplest possible setting: the H$_2$ molecule. Despite its apparent simplicity, this system already probes key challenges for VQE-based molecular dynamics, namely the need for consistent forces over long trajectories and the ability to minimize the number of expensive quantum evaluations.

\begin{figure}[htb]
  \centering
  \includegraphics[width=0.5\linewidth]{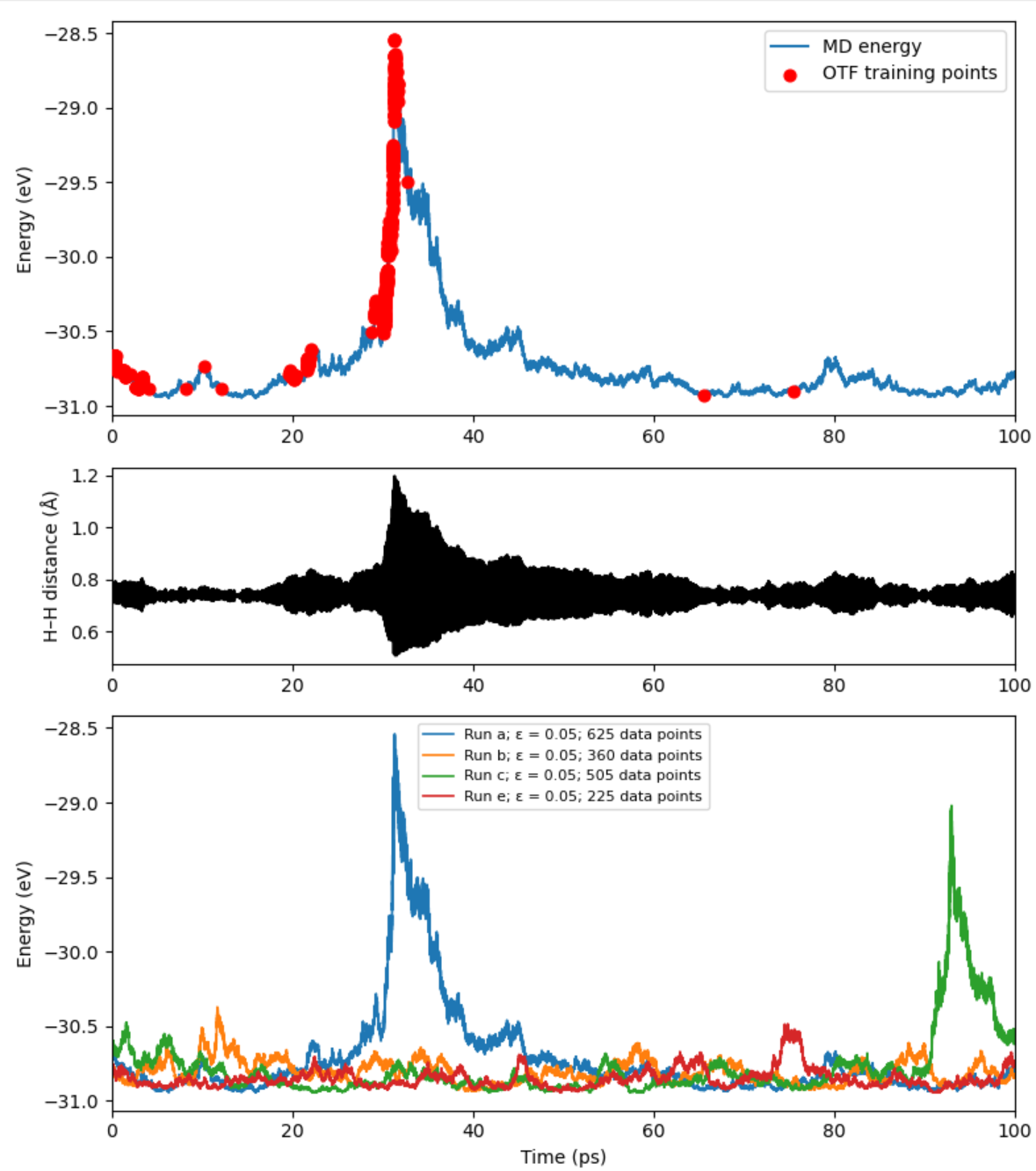}
  \caption{Molecular dynamics simulations of H$_2$ using ADAPT--VQE forces combined with on-the-fly active learning (FALCON). 
Top: Total energy along a representative trajectory; red markers indicate time steps at which additional VQE force evaluations were performed for model retraining. 
Middle: Corresponding H--H bond distance as a function of time. 
Bottom: Comparison of four independent MD runs, annotated with the total number of VQE calculations required in each case.}
  \label{fig:h2_md}
\end{figure}

Figure~\ref{fig:h2_md} shows representative molecular dynamics simulations of H$_2$ performed using ADAPT--VQE forces in combination with the on-the-fly active learning framework FALCON. The top panel displays the total energy along a representative trajectory, with red markers indicating the time steps at which additional VQE calculations were triggered to retrain the machine-learning force model. The middle panel shows the corresponding H--H bond distance as a function of time, illustrating stable vibrational motion over the full simulation window. The lower panel compares four independent molecular dynamics runs carried out with identical simulation parameters but different active-learning histories, highlighting the variability in the number of required quantum evaluations.

Across all runs, only a limited number of VQE force evaluations are required, with the majority of force calls provided by the actively trained surrogate model. The red markers in the top panel indicate that retraining events are concentrated in regions of configurational space where the bond is significantly stretched, i.e., where the local force landscape deviates most strongly from the previously learned model. Once these regions are sampled, the molecular dynamics proceeds stably without further quantum queries.

These results demonstrate that even for a minimal system, direct molecular dynamics driven purely by VQE forces would be computationally prohibitive, whereas the combination of adaptive variational quantum algorithms with on-the-fly active learning enables stable and efficient trajectories. The H$_2$ example thus serves as a proof-of-principle for force-driven quantum-classical molecular dynamics, validating the workflow before moving to more complex systems.

Having established stable active-learning–accelerated molecular dynamics for H$_2$, we next consider BeH$_2$, which introduces additional electrons, vibrational modes, and anharmonic force components, providing a significantly more demanding test of the combined ADAPT--VQE and FALCON framework.

\begin{figure}[htb]
  \centering
  \includegraphics[width=0.4\linewidth]{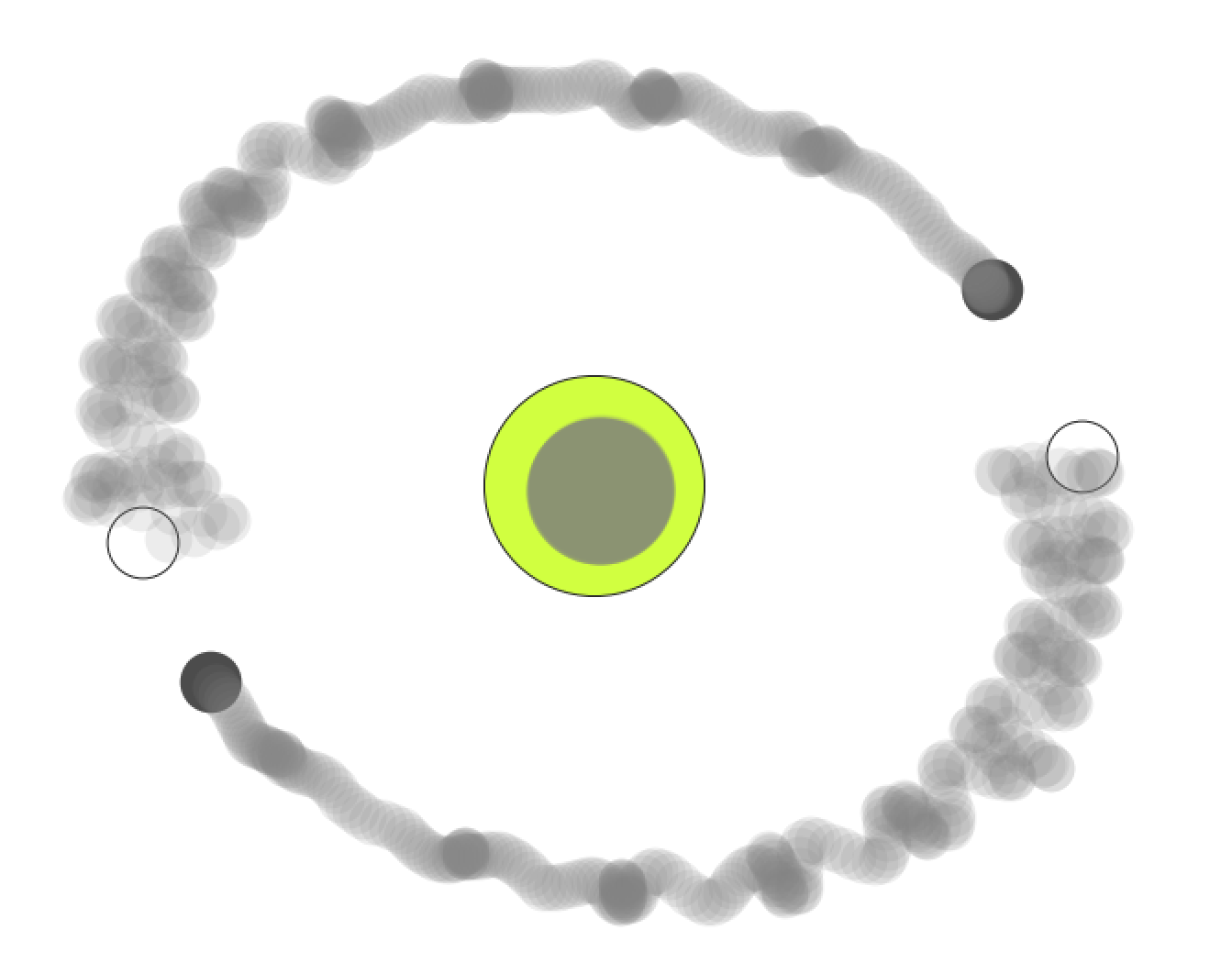}
  \caption{Trajectory of molecular dynamics run of BeH$_2$ at room temperature. Black atoms are start positions, grey are intermediate positions. In white the H atoms and green the Be atoms of the final configuration are shown, respectively.}
  \label{fig:beh2md}
\end{figure}

Using BeH$_2$ at its optimized geometry, a NVT simulation ran for 500~fs using our previously
published active-learning ML algorithm (\texttt{FALCON}) \cite{felisfalcon} to accelerate force
evaluations. A sketch of the trajectory is illustrated in figure \ref{fig:beh2md}.

The molecule undergoes a rotation at room temperature. After 250 fs simulation time a symmetrical stretching vibration can be observed.\\ 

---

In each of these progressively more complex systems, the introduction of the ADAPT-VQE ansatz was critical in achieving quantitative agreement with high-level classical theory, even when basis set limitations dominated the absolute values. Moving beyond H$_2$ to multi-electron and polyatomic systems with well-defined geometries and vibrations places the ASE–Qiskit integration squarely in the domain of realistic applications, and demonstrates that hybrid quantum–classical atomistic modelling is now feasible within familiar simulation environments.


\section{Conclusion and Outlook}

This manuscript demonstrated the integration of Qiskit Nature’s variational quantum chemistry solvers into the Atomic Simulation Environment (ASE), enabling hybrid quantum–classical workflows for atomistic simulations driven by variational quantum algorithms. By making ADAPT--VQE available as a drop-in ASE calculator, electronic structure energies, forces, and derived properties can be accessed directly within established simulation tasks such as geometry optimisation, vibrational analysis, strain evaluation, and molecular dynamics.

Within the limitations imposed by small basis sets and simulated quantum hardware, we showed that adaptive variational ansätze are essential for obtaining stable and chemically meaningful forces in multi-electron systems. In particular, ADAPT--VQE systematically outperformed fixed-ansatz VQE in reproducing CCSD reference geometries and vibrational spectra for diatomic and polyatomic molecules.

A central result of this work is the demonstration that force-driven molecular dynamics based on variational quantum algorithms becomes computationally feasible when combined with on-the-fly active learning. Using H$_2$ as a minimal test case, we showed that stable molecular dynamics trajectories can be generated with a strongly reduced number of explicit VQE force evaluations, while BeH$_2$ served as a more demanding example highlighting the robustness of the combined ADAPT--VQE and FALCON workflow in multi-electron systems.

These results indicate that hybrid quantum–classical atomistic modelling can be embedded into familiar simulation environments in a workflow-oriented manner, even when direct quantum evaluations remain expensive. Rather than relying on isolated single-point calculations, the presented approach enables sustained force-driven simulations by integrating adaptive quantum algorithms with classical acceleration techniques.

Future work will focus on incorporating analytic gradients, tailored ansätze for specific chemical classes, and improved error-mitigation strategies for real quantum hardware. Beyond standalone molecular systems, the present framework naturally lends itself to embedding and multiscale approaches, such as QM/MM or subsystem-based treatments, where quantum resources are selectively applied to chemically relevant regions while the surrounding environment is treated classically. As quantum hardware matures in qubit count and fidelity, the integration presented here establishes a practical pathway for embedding variational quantum algorithms into large-scale atomistic simulation workflows.

\section{Data Availability}
The datasets generated and analyzed for the underlying study, as well as the scripts used, are available on GitHub at \url{https://github.com/thequantumchemist/ase_quantum_vqe}. The \textbf{FALCON} code is available on GitHub at \url{https://github.com/thequantumchemist/falcon} and can be installed via \textit{pip install falcon-md}.

\section{Author Contributions}
WD did all research, coding, data analyzes and writing.

\section{Competing Interests}
The author declares no competing interests.

\section{Acknowledgement}
WD acknowledges financial support through the APF project ‘Materials on Demand’ within the ‘Humans on Mars’ Initiative funded by the Federal State of Bremen and the University of Bremen. In addition, we gratefully acknowledge funding by the Central Research Development Fund of the University of Bremen. WD thanks Rahel Weiß for testing the code.

\bibliographystyle{unsrt}       
\bibliography{lib}     
\end{document}